\documentclass[aps,pre,reprint,superscriptaddress]{revtex4-1}

\usepackage[dvips]{graphicx}

\usepackage{bm}
\usepackage{color}

\begin{document}


\title{Suppression of the Richtmyer-Meshkov instability due to a density transition layer at the interface}


\author{Takayoshi Sano}
\email{sano@ile.osaka-u.ac.jp}
\affiliation{Institute of Laser Engineering, Osaka University, Suita,
  Osaka 565-0871, Japan} 

\author{Kazuki Ishigure}
\affiliation{Institute of Laser Engineering, Osaka University, Suita,
  Osaka 565-0871, Japan} 

\author{Fransisco Cobos-Campos}
\affiliation{ETSI Industriales, Instituto de Investigaciones
  Energ{\'e}ticas and CYTEMA, Universidad de Castilla-La Mancha, 13071
  Ciudad Real, Spain}
\affiliation{{\color{black} Fluid Mechanics Group, Escuela
    Polit{\'e}cnica Superior, Universidad Carlos III de Madrid, 28911
    Legan{\'e}s (Madrid), Spain}}


\date{Jun 4, 2020; accepted for publication in Physical Review E}

\begin{abstract}
We have investigated the effects of a smooth transition layer at the contact discontinuity on the growth of the Richtmyer-Meshkov instability (RMI) by hydrodynamic numerical simulations and derived an empirical condition for the suppression of the instability.  
{\color{black}{%
The transition layer has little influence on the RMI when the thickness $L$ is narrower than the wavelength of an interface modulation $\lambda$.}}
However, if the transition layer becomes broader than $\lambda$, the perturbed velocity associated with the RMI is reduced considerably. 
{\color{black}{%
The suppression condition is interpreted as the cases that the shock transit time through the transition layer is longer than the sound crossing time of the modulation wavelength.}}
The fluctuation kinetic energy decreases as $L^{-p}$ with $p = 2.5$, which indicates that the growth velocity of the RMI decreases in proportion to $L^{-p/2}$ by the presence of the transition layer.  
This feature is found to be quite universal and appeared in a wide range of shock-interface interactions.
%
\end{abstract}


\maketitle


\section{Introduction}

Interfacial instabilities are of great importance in various plasma
phenomena in the universe and laboratory experiments
\cite{abarzhi10,nishihara10}.  
The Richtmyer-Meshkov instability (RMI) \cite{richtmyer60,meshkov69}
is one of such instabilities that occurs when a planar shock hits
a corrugated surface of the contact discontinuity.  
The RMI has been studied vigorously by the linear theory
\cite{meyer72,mikaelian93,yang94,wouchuk96,wouchuk97,vandenboomgaerde98},
nonlinear analysis \cite{zhang97,sadot98,matsuoka03,latini07,dimonte10}, and
laboratory experiments
\cite{jacobs96,brouillette99,chapman06,dimonte93,farley99,glendinning03,aglitskiy06}. 

Turbulent mixing excited by the RMI often plays a crucial role
associated with plasma explosions in astrophysical objects
\cite{makee77} and the implosion in inertial confinement fusion
\cite{atzeni04,betti16}. 
Interaction of supernova shocks and inhomogeneous interstellar matters
is one of the promising sites of the RMI, which could contribute to
the origin of the interstellar turbulence \cite{inoue09} as well as
the amplification of magnetic fields \cite{sano12}. 
The RMI is recognized as one of the severe obstacles to prevent the
ideal implosion in laser fusion plasmas \cite{atzeni04,betti16}. 
Drastic symmetry reduction results in inadequate energy gain at the end of
the process.
Therefore, the mitigation mechanisms of the RMI are paid attention
intensely in this field.

There are several effects proposed to stabilize the RMI.
The vorticity deposited at the interface just after the incident shock
refraction is the driving source of the RMI growth, while the
vorticity left in the bulk of the fluids
has been proved to be a physical agent that decreases the 
growth of the contact surface ripple
\cite{coboscampos16,coboscampos17}.
However, the effect of the bulk vorticity becomes significant only
when the shock is sufficiently strong, or the compression is high
enough.   
{\color{black}{%
For the RMI in plasmas,}}
a strong magnetic field can suppress the growth of the RMI when the
Alfv{\'e}n (Mach) number, which is the ratio of the linear growth velocity to
the Alfv{\'e}n speed, is less than unity
\cite{samtaney03,wheatley05,sano13}. 
However, if the direction of the magnetic field is parallel to the
interface but perpendicular to the wavevector of the surface
modulation, the Lorentz force hardly works on the RMI. 
Then, the suppression by the magnetic field in three-dimensional
geometry would be difficult so as in the case of the Rayleigh-Taylor
instability \cite{chandrasekhar61,stone07}.

In this paper, we focus on the effect of a density transition layer at
the interface for the suppression of the RMI.
It is well known that the smooth density structure of the interface
affects the unstable growth of surface fluctuations. 
For example, the density stratification at the shear layer of the
velocity stabilizes the Kelvin-Helmholtz instability
\cite{chandrasekhar61}.
The stability condition is given by the Richardson number, which is a
function of the density gradient. 
The Rayleigh-Taylor instability is also affected by the density stratification \cite{munro88,atzeni04}.
The growth rate of the instability decreases dramatically if the scale
length of the density structure is longer than the wavelength of the 
Rayleigh-Taylor mode. 
As for the RMI, the transition-layer effects have not been much
investigated because of the difficulty of the analytical approach
\cite{mikaelian85}. 

The density transition layer is naturally formed in astrophysical
objects and laser plasmas.
The molecular clouds in the interstellar medium are modeled by an isothermal self-gravitating sphere, which is the so-called Bonnor-Ebert sphere, where a flat high-density core surrounded by a power-law envelope.  
The smooth distribution of the density should affect the stability at the shock interaction \cite{klein94,nakamura06,falle17}.
In laboratory plasmas, the density distribution in the laser ablation
layer is verified to mitigate the turbulent-mixing caused by the
ablative Rayleigh-Taylor instability \cite{takabe85,betti98}.  
The exponential density distribution is usually assumed at the
material interface in this case.
Thus, the RMI with a smooth layer of the density transition would have
numerous critical applications. 

The goal of this paper is to obtain the suppression condition of RMI
due to the existence of the density transition layer by using
nonlinear hydrodynamic simulations. 
The analytical treatment of the RMI is not straightforward when the
interface has a non-uniform density profile.
In that case, numerical simulations are a powerful tool as the first
step to examine such complicated situations and to extract the essence
of the physical basis empirically.

The outline of this paper is as follows.  
In Sec.~\ref{sec2}, the basic equations, initial conditions, and numerical methods are described.  
Various simulation results are shown in Sec.~\ref{sec3} to reveal the influence of the transition layer on the growth of the RMI.
In Sec.~\ref{sec4}, the physical interpretation of our findings is discussed. 
Then, the suppression condition of the RMI in terms of the thickness of the transition layer is derived.
We also remark on an application of our results to laboratory laser plasmas. 
Finally, the conclusions are summarized in Sec.~\ref{sec5}.

\section{Numerical Method \label{sec2}}

To study the nonlinear evolutions of the RMI, the following system
equations for inviscid fluids are solved; 
\begin{eqnarray}
\frac{\partial \rho}{\partial t} + {\mbox{\boldmath{$\nabla$}}} \cdot
\left( \rho {\mbox{\boldmath{$v$}}} \right) = 0 \;, 
\label{eq:con}
\\
\frac{\partial (\rho {\mbox{\boldmath{$v$}}})}{\partial t} +
{\mbox{\boldmath{$\nabla$}}} \cdot
\left[ P {\mbox{\boldmath{$I$}}}
+ \rho {\mbox{\boldmath{$v$}}}
{\mbox{\boldmath{$v$}}} \right] = 0 \;,
\\
\frac{\partial e}{\partial t} + {\mbox{\boldmath{$\nabla$}}} \cdot
\left[ \left( e + P \right) 
{\mbox{\boldmath{$v$}}} 
\right] = 0 \;,
\label{eq:ene}
\end{eqnarray}
where $\rho$ and {\mbox{\boldmath{$v$}}} are the mass density and
velocity, respectively, and $e$ is the total energy
    density per unit volume, $e = P/(\gamma -1) + \rho v^2/2$.
{\color{black}
The equation of state for the ideal gas is used with the isentropic
exponent $\gamma$.}


\begin{figure}
\includegraphics[scale=0.85,clip]{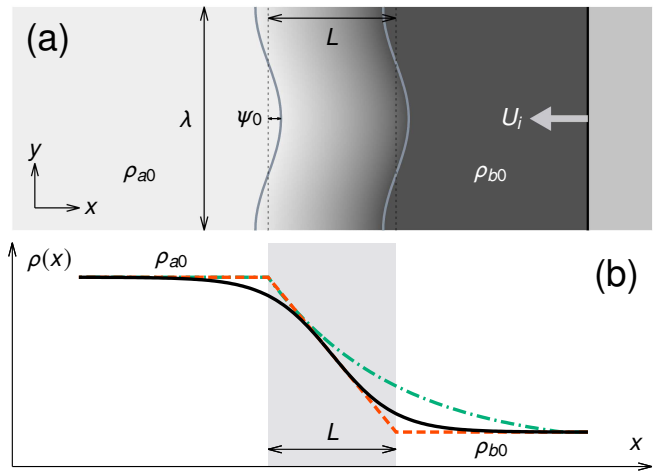}%
\caption{
(a)
Initial setup of the single-mode analysis for the RMI.
A planar shock hits a corrugated interface between the fluid
``a'' and ``b'', where the wavelength of the surface modulation is
assumed to be $\lambda$. 
{\color{black}
The characteristic quantities of this system are the incident shock
velocity $U_i$, the density jump at the interface $\rho_{a0} / \rho_{b0}$,
and the corrugation amplitude $\psi_0$.}
In this analysis, the density transition layer with a thickness of $L$
is considered. 
(b)
Density distributions of the transition layer adopted in this
analysis, which are the $\tanh$-type (black), linear-type (red), and
$\exp$-type (green). 
The transition layer is located at $-L/2 \le x \le L/2$ initially. 
\label{fig1}}
\end{figure}

We adopt a single-mode setup for our numerical analysis, which is
illustrated by Fig.~\ref{fig1}(a). 
Two fluids with different densities, $\rho_{a0}$ and $\rho_{b0}$,
are separated by a corrugated interface located at $x = 0$.
A planar shock propagating through the fluid ``b'' hits the corrugated
interface at $t = 0$.
Here the $x$- and $y$-axis are set to be perpendicular and parallel to
the shock surface.  
The incident shock velocity is $-U_i \hat{x}$, where $\hat{x}$ is a
unit vector.
{\color{black}{Both the fluids}} 
are stationary $\bm{v} = 0$ and have a uniform pressure $P_0$
before the shock passage. 
The sonic Mach number of the incident shock is defined as $M = U_i
/c_{b0}$ where $c_{b0} = (\gamma_b P_0/ \rho_{b0})^{1/2}$ is the sound
speed of the fluid ``b''. 
The physical quantities in the post-shocked region behind the incident
shock are calculated from the Rankine-Hugoniot conditions.
The interface has an initial corrugation of a sinusoidal form, $x =
\psi_0 \cos ( k y )$, where $\psi_0$ is the corrugation amplitude, $k =
2 \pi / \lambda$ is the perturbation wavenumber, and $\lambda$ is the
wavelength.

{\color{black}
Six non-dimensional parameters characterize the initial
configuration of the nonlinear single-mode analysis with a transition layer.  
The Mach number $M$ parameterizes the incident shock velocity.  
The contact discontinuity is expressed by the density jump
$\rho_{a0} / \rho_{b0}$ and the ratio of the corrugation amplitude
to the wavelength $\psi_0 / \lambda$.
The compressibility indicated by the
isentropic exponents $\gamma_a$ and $\gamma_b$ of each fluid is also an
essential element of this system.
Besides, we introduce a density transition layer with a
finite width of $L$, which provides an additional parameter of $L /
\lambda$. 
In this analysis, we assume a constant value of $\gamma_a = \gamma_b =
5/3$ for simplicity.
Thus, the other four parameters ($M$, $\rho_{a0} / \rho_{b0}$, $\psi_0
/ \lambda$, and $L / \lambda$) are considered in the following.
}

Various shapes of the density distribution in the layer could be
plausible according to circumstances.
As a typical function, we take a hyperbolic-tangent function expressed 
as  
\begin{equation}
  \rho (x) = \frac{\rho_{a0} + \rho_{b0}}{2} \left[
  1 - A \tanh \left( \frac{2 x}{L} \right)  \right] \;,
  \label{eq:tanh}
\end{equation}
where $A = (\rho_{a0} - \rho_{b0}) /  (\rho_{a0} + \rho_{b0})$ is the
Atwood number.
Linear and exponential distributions are also examined for comparison,
which are given by  
\begin{equation}
  \rho (x) =
  \left\{
  \begin{array}{lcl}
    \rho_{a0} & & \displaystyle \left(x \leq  -\frac{L}2 \right) \\
  \displaystyle \frac{\rho_{a0} + \rho_{b0}}{2} \left[
   1 - A \left( \frac{2 x}{L} \right)  \right] & &
    \displaystyle \left( |x| < \frac{L}2 \right) \\
    \rho_{b0} & & \displaystyle \left( x \geq \frac{L}2 \right) 
  \end{array} \right.
  \label{eq:lin}  
\end{equation}
and 
\begin{equation}
  \rho (x) =
  \left\{
  \begin{array}{lcl}
    \rho_{a0} & & \displaystyle \left( x \leq - \frac{L}2 \right) \\
    \displaystyle \max \left\{ \rho_{a0} e^{
    - \frac1{L} \left| x + \frac{L}2 \right|} , \rho_{b0} \right\}
    & & \displaystyle \left( x > - \frac{L}2 \right) 
  \end{array} \right. \;.
    \label{eq:exp}  
\end{equation}
While the thickness of the transition layer is well defined by $L$ in
the hyperbolic-tangent and linear distributions, the effective thickness in
the exponential-type depends on the combination of the scale length $L$ and
density jump $\rho_{a0} / \rho_{b0}$. 

There are several formulas suggested evaluating the linear growth
velocity of the RMI theoretically. 
The linear growth with time, not exponential, is one of the unique
characteristics of the RMI. 
Another feature of the RMI is that it occurs in both cases of
light-to-heavy ($\rho_{a0} / \rho_{b0} > 1$) and heavy-to-light
($\rho_{a0} / \rho_{b0} < 1$) configurations. 

\citet{richtmyer60} was the first to study the problem of a planar
shock crossing the corrugated boundary between two fluids, and
proposed a generalization of the Rayleigh-Taylor formula as the growth
velocity 
$\partial \psi /  \partial t = k v^{\ast} A^{\ast} \psi^{\ast}_0$,
where $v^{\ast}$ and $\psi^{\ast}_0 = \psi_0 ( 1 - v^{\ast} / U_i )$
are the zero-order velocity and the amplitude of the contact surface
just after the shock passage, respectively.
The Atwood number $A^{\ast} = ( \rho_{a}^{\ast} - \rho_{b}^{\ast} ) /
( \rho_{a}^{\ast} + \rho_{b}^{\ast} )$ is defined by the densities at
both sides of the post-shocked interface.
Then, \citet{meyer72} observed that the Ricthmyer prescription should be
modified using an averaged value between the pre- and post-shocked
interface amplitude, i.e., $\partial \psi / \partial t = k v^{\ast}
A^{\ast} ( \psi_0 + \psi^{\ast}_0 ) / 2$, in order to obtain agreement
between the numerical solution and the linear theory. 
A similar heuristic approach was also proposed by
\citet{vandenboomgaerde98}.  
Unfortunately, these empirical prescriptions are likely to fail for
high compressions \cite{vandenboomgaerde98}.

Further linear theories of the RMI have been done in the form of series
expansions in terms of inverse powers of the Laplace variable
\cite{fraley86}, in powers of the time \cite{velikovich96}, or in terms of
the Bessel functions \cite{wouchuk96,wouchuk97,coboscampos16,coboscampos17}.
In particular, the asymptotic growth velocity for both shock- and
rarefaction-reflected cases is calculated with the following
expression derived by \citet{wouchuk97}:
\begin{equation}
  v_{\rm wn} =
  \frac{- \rho_{a}^{\ast} \delta v_{ya}^{\ast}
        + \rho_{b}^{\ast} \delta v_{yb}^{\ast}}
       {\rho_{a}^{\ast} + \rho_{b}^{\ast}} +
  \frac{\rho_{a}^{\ast} F_a - \rho_{b}^{\ast} F_b }
       {\rho_{a}^{\ast} + \rho_{b}^{\ast}} \;, 
\label{eq:dvi}
\end{equation}
where $\delta v_{ya}^{\ast}$ and $\delta v_{yb}^{\ast}$ are the initial
tangential velocities at both sides of the contact surface.
The quantities $F_a$ and $F_b$ represent the sonic interaction between
the contact surface and the transmitted and reflected wavefront,
respectively, which are proportional to the amount of vorticity left
behind the wavefronts in the bulk of each fluid.
For the case when a rarefaction is reflected, no vorticity is created
in the expanded fluid, i.e., $F_b=0$. 

The Wouchuk-Nishihara (WN) formula is rigorously deduced from
linearized two-dimensional Euler equations after two wavefronts have
separated away from the interface.
The growth velocity given by Eq.~(\ref{eq:dvi}) is exact within the
limits of linear theory and inviscid flow. 
It is valid for any initial configuration, and every element can be
analytically calculated from the pre-shocked parameters
\cite{wouchuk01a,wouchuk01b,coboscampos16,coboscampos17}.
The first term of the right-hand side of Eq. (\ref{eq:dvi}) is due
to the instantaneous deposition of the vorticity at the interface just
after the shock interaction.
On the other hand, the second term represents the interaction between
the contact surface and the wavefronts. 
It becomes non-negligible for stronger shocks or highly compressible
fluids, and typically has the opposite sign to the first term.
The negative growth velocity stands for the phase reversal that could
occur the rarefaction-reflected cases. 
Throughout our analysis, the WN formula $v_{\rm wn}$ is 
used as the typical velocity of the RMI growth for a given set of
the parameters ($M$, $\rho_{a0} / \rho_{b0}$, and $\psi_0 / \lambda$).

We solve the system equations (\ref{eq:con})-(\ref{eq:ene}) in
two-dimension ($x$, $y$) {\color{black}{in the Cartesian coordinate system}}
by using a conservative Godunov scheme with the second-order accuracy in
space and time \cite{vanleer79,sano98}.
The exact solutions of the Riemann problem at each grid boundary are
used in the flux calculations for time integration of the variables
\cite{colella84}. 
The scheme includes an additional numerical diffusion in the
direction tangential to the shock surface in order to care for the
carbuncle instability \cite{hanawa08}.   
A periodic boundary condition is used in the
$y$-direction, and an 
outflow boundary condition is adopted in the $x$-direction.  
The size of the computational box in the $y$-direction
is always set to be $L_y = \lambda$. 
The choice of $L_x$, on the other hand, depends on the
initial parameters. 
The $x$-length is taken to be sufficiently extensive so
that both of 
the transmitted shock and reflected shock (or rarefaction) never reach
the edge of the computational domain in all the runs.  

Most of the calculations are performed with a standard resolution of
$\Delta_x$ = $\Delta_y$ = $\lambda/256$ unless otherwise stated. 
The physical quantities are normalized by the initial density and
sound speed of the fluid ``b'', $\rho_{b0} = 1$ and $c_{b0} =
1$, and the wavelength of the surface modulation $\lambda = 1$. 
The sound crossing time of the wavelength is also unity in our
normalization, $\lambda / c_{b0} = 1$. 

\section{Numerical Results \label{sec3}}


\begin{figure*}
\includegraphics[scale=0.85,clip]{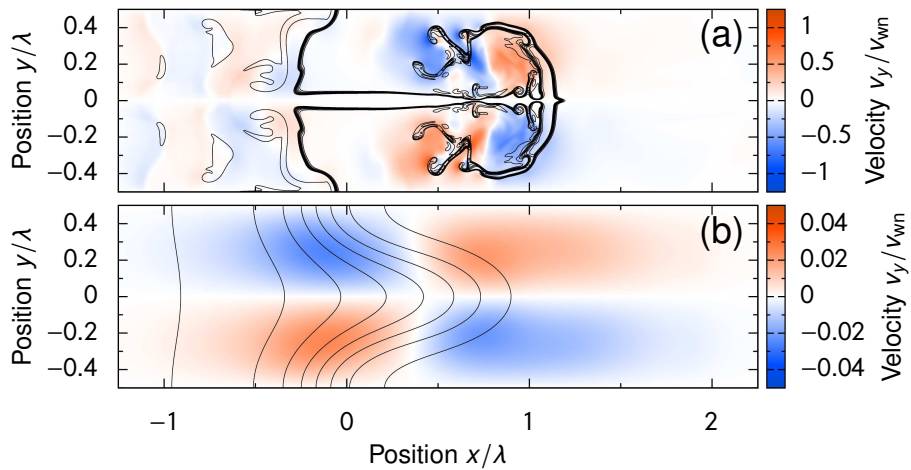}%
\caption{
Spatial distributions of the density $\rho$ and the tangential
velocity $v_y$ at the nonlinear regime of the RMI growth $k v_{\rm wn}
t = 15$ for the cases of (a) $L / \lambda = 0.03$ and (b) $L / \lambda
= 3$.  
The other parameters are identical for both cases, which are $M = 2$,
$\rho_{a0}/\rho_{b0} = 10$, and $\psi_0 / \lambda = 0.1$. 
The $\tanh$-type transition layer is assumed.
The density distribution is depicted by the contour curves at the
levels from $\rho / \rho_{b0} = 6$ to 27. 
The color map denotes the tangential velocity normalized by the growth
velocity of the WN model $v_y / v_{\rm wn}$. 
The color-bar range of the velocity in (b) is $1/25$ of that in (a). 
\label{fig2}}
\end{figure*}

First, we look at the difference in the density distribution caused by
the existence of a transition layer.  
The growth of the RMI is usually recognized by the mushroom-shaped
elongation of the density interface.  
Figure~\ref{fig2} shows the contour lines of the density at the nonlinear regime
of the RMI growth.  
The horizontal axis $x$ is converted to a frame moving with the
contact-discontinuity velocity $v^{\ast}$ after the shock interaction. 
The initial parameters in this fiducial run are the Mach number $M=2$,
the density jump $\rho_{a0}/\rho_{b0} = 10$, and the modulation
amplitude $\psi_0 / \lambda = 0.1$. 
The function of the density transition layer is the hyperbolic tangent
given by Eq.~(\ref{eq:tanh}). 
The snapshot is taken at $k v_{\rm wn} t = 15$, where $( k v_{\rm wn
})^{-1}$ is the unit timescale of the RMI.
The linear growth velocity of the WN model $v_{\rm wn}$ is evaluated
from Eq.~(\ref{eq:dvi}).
For the fiducial parameters, $v_{\rm wn} / c_{b0} = 0.20651$, so that
the RMI timescale corresponds to slightly shorter than the initial
sound crossing time, $( k v_{\rm wn} )^{-1} \sim 0.77 ( \lambda /
c_{b0} )$.  

For a narrow transition case of $L / \lambda = 0.03$
[Fig.~\ref{fig2}(a)], 
the RMI growth is nearly identical to the case with a sharp boundary
case ($L = 0$).
The width of the mixing layer defined
from the spike top to bubble bottom exceeds the modulation wavelength
of $\lambda$.
On the other hand, the deformation of the interface is significantly
reduced when the transition layer becomes comparable to $\lambda$.  
Figure~\ref{fig2}(b) shows the density contours for a case of $L /
\lambda = 3$, in which the other parameters are the same as those of
Fig.~\ref{fig2}(a).  
Although the location of the interface cannot be defined uniquely for
this case, the density contours are rather smooth compared to those in
Fig.~\ref{fig2}(a).  
The fluctuation amplitude of the contour lines is at most a few times
larger than the initial corrugation amplitude of $\psi_0 / \lambda =
0.1$.  
Thus, the transition layer indeed mitigates the growth of the RMI. 
The enhancement of the modulation in the density structure is severely
suppressed. 

Because the RMI growth is tightly connected to the tangential velocity
induced by shock interaction with a corrugated interface, we
focus on the $y$-component of the perturbed velocity in our
simulations.   
Note that $v_y$ is nothing everywhere before the shock passage since
we consider homogeneous initial flow. 
In Figs.~\ref{fig2}(a) and \ref{fig2}(b), the tangential velocity
$v_y$ normalized by $v_{\rm wn}$ is depicted by colors for each case.

The tangential velocity is of the order of $v_{\rm wn}$ when $L /
\lambda = 0.03$ [Fig.~\ref{fig2}(a)].
At the time of the snapshot $k v_{\rm wn} t = 15$, the fastest
velocity is localized at the roll-up region of the mushroom shape.
The mixing-layer width 
due to the RMI motions
is still growing even at this nonlinear phase. 
By contrast, the generation of $v_y$ is weakened by more than an order
of magnitude in the broad transition case of $L/ \lambda = 3$
[Fig.~\ref{fig2}(b)]. 
The color range of the tangential velocity in Fig.~\ref{fig2}(b) is
about $1/25$ of that in Fig.~\ref{fig2}(a). 
Weak tangential shear is deposited in the middle of the transition
layer, where the most considerable distortion of the density contour
is observed.
The location of the maximum vorticity would be related to the largest
gradient of the density.


\begin{figure*}
\includegraphics[scale=0.85,clip]{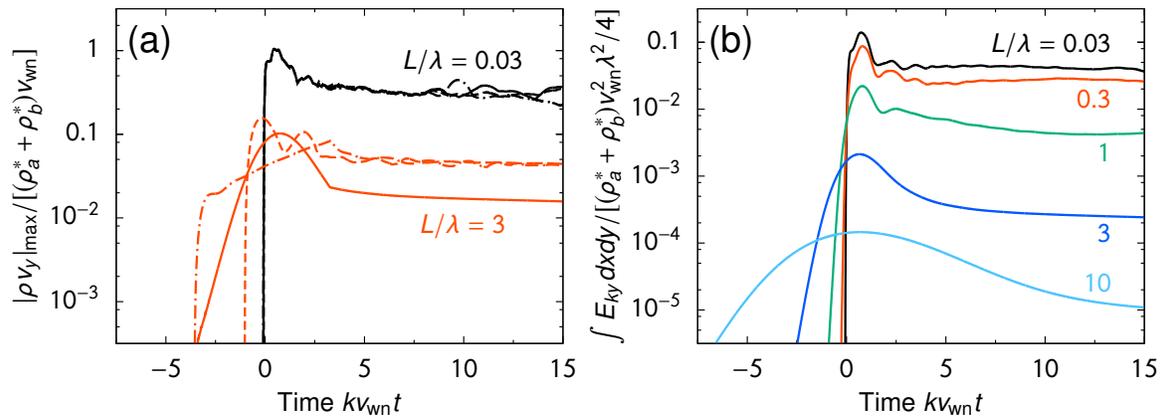}%
\caption{
(a)
Time evolutions of the maximum tangential momentum $|\rho v_y|$ in the
entire area of the computational domain for the cases of $L/\lambda =
0.03$ (black) and $L / \lambda = 3$ (red). 
The other parameters are identical to the fiducial runs shown in
Fig.~\ref{fig2}.  
The dependence on the density function is also indicated in this
figure by the different line types. 
The results of the $\tanh$-type, liner-type, and $\exp$-type are depicted
by the solid, dashed, and dot-dashed curves, respectively.
The time in the horizontal axis is given in the unit of the RMI
timescale $( k v_{\rm wn} )^{-1}$.  
(b)
Time histories of
the kinetic energy
defined by the tangential
velocity $E_{ky} \equiv \rho v_y^2 / 2$ integrated over the entire
domain. 
The thickness of the $\tanh$-type transition layer is labeled for each
curve.  
The other parameters are the same as those of the runs shown in (a). 
\label{fig3}}
\end{figure*}

\begin{figure*}
\includegraphics[scale=0.85,clip]{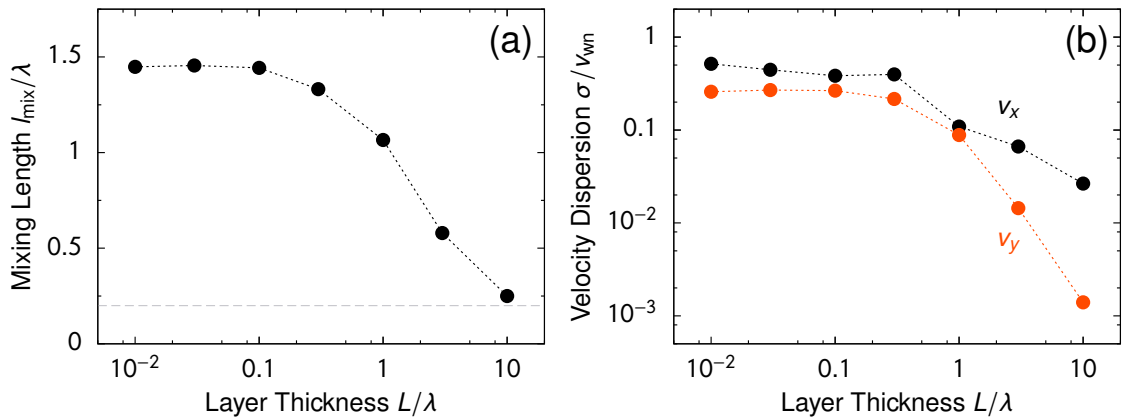}%
\caption{
(a)
Dependence of the mixing length $l_{\rm mix}$ caused by the RMI growth
on the thickness of the transition layer.
The mixing length is defined by the distance from the spike top to the
bubble bottom, which is evaluated through the tracer particle positions
at $k v_{\rm wn} t = 15$.
The simulation parameters are the same as those in the fiducial run
except for $L$.
The gray dashed line denotes the initial amplitude of $2 \psi_0 /
\lambda = 0.2$.
(b)
Standard deviation $\sigma$ of the interface velocities $v_x$ and $v_y$ in
terms of the layer thickness $L$.
The interface velocities are taken at $k v_{\rm wn} t = 15$ in the
same runs as in (a).
\label{fig8}}
\end{figure*}

The tangential velocity is a good indicator of the activity of the RMI. 
The time evolutions of the maximum value of $| \rho v_y|$ for two cases
in Fig.~\ref{fig2} are shown by the solid curves in Fig.~\ref{fig3}. 
The maximum momentum is divided by $(\rho_{a}^{\ast} +
\rho_{b}^{\ast}) v_{\rm wn}$ so that the vertical axis approximately
indicates the effective growth velocity relative to the original
$v_{\rm wn}$ of the WN model.  
When the transition layer is narrow ($L / \lambda = 0.03$), the growth
velocity appears instantaneously at $t = 0$ and keeps nearly constant
around $\sim v_{\rm wn}$. 
However, the growth velocity in the broader transition case ($L /
\lambda = 3$) increases gradually during the shock travels in the
smooth density distribution. 
The peak value is much lower than the linear-theory prediction for
the discontinuous case. 
Since the density changes continuously in the transition layer,
the shock pressure at the downstream, or the shock strength, is
weakened compared to the discontinuous case. 
Such an effectively weaker shock causes a significant reduction of
the growth velocity of RMI in the linear phase \cite{wouchuk01a,wouchuk01b}.    

The suppression effect due to non-zero $L$ seems to have little
dependence on the functional form of the density in the transition
layer.   
The time histories of the maximum tangential momentum in the linear-type
[Eq.~(\ref{eq:lin})] and exponential-type [Eq.~(\ref{eq:exp})]
distribution are also plotted in Fig.~\ref{fig3}(a) by the dashed and
dot-dashed curves, respectively.   
The difference in the density function is negligible if the layer is
much narrower than the modulation wavelength.
Huge decrease of the growth velocity $v_y$ is observed in all the cases
of $L / \lambda = 3$,
where the peak and asymptotic velocities are quite similar among the
different function cases.

The fluctuation kinetic energy
at the nonlinear regime of RMI may be a
useful quantity to evaluate the suppression effect by the transition
layer. 
Figure~\ref{fig3}(b) shows the evolutions of the perturbed
kinetic
energy defined by $E_{ky} \equiv \rho v_y^2 / 2$ integrated over the entire
region. 
The initial parameters are the same as in Fig.~\ref{fig2} except for
the thickness of the transition layer $L$. 
As can be seen,
the fluctuation kinetic energy
decreases drastically if
$L/\lambda \gtrsim 1$. 
When $L/\lambda = 10$, for example, the integrated $E_{ky}$ is reduced
by about three orders of magnitude compared to the sharp
transition case $L / \lambda \sim 0$.
In the $x$-direction, it is hard to define the perturbed velocity
because the unperturbed distribution is also time-dependent. 
However, the $x$-component of
the perturbed velocity
must be comparable to that in the $y$-direction in the RMI motions.  
Thus we believe that Figure~\ref{fig3}(b) is representing
the fluctuation kinetic energy
driven by the RMI properly.

The $L$ dependence shown by Figs.~\ref{fig3}(a) and
\ref{fig3}(b) are obtained by the simulations with the resolution of
$\Delta_x = \Delta_y = \lambda / 512$. 
These results are found to be unaffected by the numerical resolution,
which is confirmed by identical calculations with different grid sizes
of $\lambda / 1024$ and $\lambda / 256$. 
For the case of $L / \lambda = 0.01$, the transition layer is captured
by only five grids, so that its result would be regarded as that for
the discontinuous case. 

Lagrangian tracer particles are often used to pursue the evolution of the interface shape and velocity distribution for the case of a sharp density jump.
Here we apply this method even for the cases with a finite transition layer.
The tracer particles are set initially at the center of the transition
layer, that is, along a line given by $x = \psi_0 \cos (k y)$ for all cases.
Mixing length $l_{\rm mix}$ is calculated from the difference between the maximum and minimum values of the $x$-coordinate among these particles.
Figure~\ref{fig8}(a) shows the characteristic mixing length by the RMI growth at $k v_{\rm wn} t = 15$ for various runs with different thickness of the transition layer.
The modulation amplitude at the nonlinear stage of the RMI has an apparent dependence on $L$.
For the case of $L/\lambda = 10$, the mixing length is almost the same as the initial modulation amplitude.
This fact indicates the severe suppression of the RMI due to the
transition layer, which is consistent with the result shown in
Figs.~\ref{fig2} and \ref{fig3}.

The perturbed velocity of the interface is inferred from the tracer particle velocities.
We can evaluate the standard deviation $\sigma$ of the interface velocity (or the velocity dispersion), which is shown in Fig.~\ref{fig8}(b).
Both components of $v_x$ and $v_y$ exhibit the same trend of the $L$-dependence, as seen in Fig.~\ref{fig8}(a).
Since the average velocity of $v_y$ is zero, the velocity dispersion $\sigma$ is identical to the root-mean-square of $v_y$.
If the thickness of the transition layer is negligible compared to the modulation wavelength, the perturbed velocity is comparable to the growth velocity of the WN model.
Although the velocity dispersion of $v_x$ in this limit is slightly larger than that of $v_y$, the difference is no more than double.
On the other hand, the unperturbed profile of $v_x$ in the transition layer depends on the position $x$. 
Thus, the velocity dispersion may tend to be larger than the perturbed component alone, as the transition layer becomes thick.
This is another reason why we concentrate the $y$-component of the perturbed velocity.


\begin{figure}
\includegraphics[scale=0.85,clip]{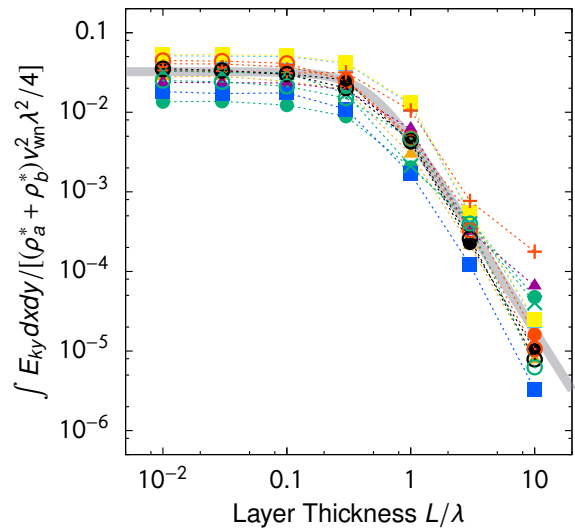}%
\caption{
Dependence of the integrated
kinetic
energy $\int E_{ky} dxdy$
measured at the nonlinear regime $k | v_{\rm wn} | t = 15$ on the
thickness of the transition layer $L$. 
Various parameter runs listed in Table~\ref{tab1} are plotted with
different marks. 
The meaning of each mark is described in the last column of
Table~\ref{tab1}. 
The gray thick curve is the fitted function of all the data, which is
proportional to $[ 1 + ( qL / \lambda )^{p} ]^{-1}$ with $q = 2.11$
and $p = 2.46$. 
\label{fig4}}
\end{figure}


\begin{table*}
\caption{
A list of the initial conditions for the simulations shown in
Fig.~\ref{fig4}. 
The key parameters are the Mach number of the incident shock $M$, the
density jump at the contact discontinuity $\rho_{a0} / \rho_{b0}$, and
the modulation amplitude relative to the wavelength of $\psi_0 /
\lambda$. 
Three types of the transition layer ($\tanh$, linear, and $\exp$) are
considered. 
The linear growth velocity $v_{\rm wn}$ of the RMI is calculated based
on the WN model.  
The obtained
kinetic
energy is fitted by $[ 1 + ( q L / \lambda
  )^{p} ]^{-1}$ as a function of $L/\lambda$, and the fitted results
for $q$ and $p$ are listed for each parameter set in the table.  
A numerical factor $\zeta$ for each run is calculated
by solving appropriate Riemann problems.
The last column is the mark of the plot in Fig.~\ref{fig4} for each
case.  
\label{tab1}}
  \begin{tabular}{cccccccccc}
  \hline \hline
  $M$ &
  $\displaystyle{\frac{\rho_{a0}}{\rho_{b0}}}$ &
  $\displaystyle{\frac{\psi_0}{\lambda}}$ &
  transition &
  $\displaystyle{\frac{v_{\rm wn}}{k \psi_0 U_i}}$ &
  $\displaystyle{\frac{v_{\rm wn}}{c_{b0}}}$ &
  $q$ &
  $p$ & $\zeta$ & mark \\ \hline
  2   & 10   & 0.1  & tanh   & 0.16433  & 0.20651  & 2.08 & 2.66 & 1.8 &
  black filled circle \\
  2   & 10   & 0.1  & linear & 0.16433  & 0.20651  & 1.87 & 1.89 & 1.8 &
  red plus \\
  2   & 10   & 0.1  & exp    & 0.16433  & 0.20651  & 3.90 & 1.84 & 1.8 &
  green cross \\
  1.2 & 10   & 0.1  & tanh   & 0.082788 & 0.062421 & 2.19 & 2.72 & 1.6 &
  blue square \\
  10  & 10   & 0.1  & tanh   & 0.18589  & 1.1680   & 1.58 & 2.79 & 2.0 &
  cyan square \\
  100 & 10   & 0.1  & tanh   & 0.18639  & 11.711   & 1.56 & 2.80 & 2.0 &
  yellow square \\
  2   & 3    & 0.1  & tanh   & 0.13365  & 0.16795  & 2.01 & 2.53 & 1.6 &
  red filled circle \\
  2   & 100  & 0.1  & tanh   & 0.081918 & 0.10294  & 2.56 & 1.73 & 2.0 &
  green filled circle \\
  2   & 0.3  & 0.1  & tanh   & $-0.20979$ & $-0.26364$ & 2.17 & 2.65 &
  1.1 &
  red open circle \\
  2   & 0.1  & 0.1  & tanh   & $-0.35302$ & $-0.44362$ & 1.92 & 2.77 &
  0.80 &
  black open circle \\
  2   & 0.01 & 0.1  & tanh   & $-0.44584$ & $-0.56026$ & 1.39 & 3.01 &
  0.34 &
  green open circle \\
  2   & 10   & 0.03 & tanh   & 0.16433  & 0.061953 & 2.16 & 2.63 & 1.8 &
  orange triangle \\
  2   & 10   & 0.3  & tanh   & 0.16433  & 0.61953  & 2.09 & 2.03 & 1.8 &
  purple triangle \\
\hline \hline
\end{tabular}
\end{table*}

It is found that the dependence of the RMI suppression on $L/\lambda$
is quite robust and valid for a wide range of the initial parameters. 
The fluctuation kinetic energy
measured at $k | v_{\rm wn} | t = 15$ for
various cases listed in Table~\ref{tab1} are shown all together in
Fig.~\ref{fig4}. 
Normalization of the
kinetic
energy in this diagram is to divide by
$( \rho_{a}^{\ast} + \rho_{b}^{\ast} ) v_{\rm wn}^2 \lambda^2 / 4$ that is
proportional to $v_{\rm wn}^2$. 
It should be noted that the dimensionless linear growth velocity
$v_{\rm wn}/(k \psi_0 U_i)$ is determined by $M$, $\rho_{a0}/\rho_{b0}$,
and $\gamma_a=\gamma_b=5/3$ in our system.
For example, a higher Mach number gives a faster growth velocity so
that the normalization factor is larger for the higher $M$ case.
The growth velocity $v_{\rm wn}$ for each case is also listed in Table~\ref{tab1}. 

All the data points in Fig.~\ref{fig4} exhibit a single
trend on $L / \lambda$, even though the vast parameter range of many
orders of magnitude are examined here.
The fiducial set of the parameters are chosen as $M=2$, $\rho_{a0} /
\rho_{b0} = 10$, $\psi_0 / \lambda = 0.1$, and $\tanh$-type function.  
Then we examine the dependence of the transition function (linear- and
$\exp$-type), the Mach number ($M=1.2$, 10, and 100), the density jump
for shock-reflected cases ($\rho_{a0} / \rho_{b0} = 3$ and 100) and for
rarefaction-reflected cases ($\rho_{a0} / \rho_{b0} = 0.3$, 0.1, and
0.01), and the modulation amplitude ($\psi_0 / \lambda = 0.03$ and
0.3).  
For each parameter case, we perform seven runs with different widths of
the transition layer in a range from $L / \lambda = 0.01$ to 10 to
identify the dependence. 

When $L/\lambda \lesssim 1$,
the fluctuation kinetic energy
is flat and almost
identical to that in the discontinuous limit $L \rightarrow 0$. 
On the other hand, if $L/\lambda$ becomes larger than unity, the
kinetic
energy decreases with a power law. 
The dependence could be fitted by a function proportional to $[
1 + (q L / \lambda)^{p} ]^{-1}$ with two fitting parameters $q$ and $p$. 
The fitted results for $q$ and $p$ are listed in Table~\ref{tab1} for
each parameter case. 
The average values of all the cases are $q = 2.11 \pm 0.35$ and $p =
2.46 \pm 0.17$, which is drawn by the gray thick curve in
Fig.~\ref{fig4}.
Thus, the influence of the transition layer begins
to appear when $L \gtrsim \lambda / 2$.
In the limit of $L / \lambda \gg 1$,
the fluctuation kinetic
drops in
proportion to $L^{-5/2}$, which means the growth velocity has a
power-law dependence of $L^{-5/4}$ approximately. 

\section{Discussion \label{sec4}}

Our numerical results suggest that the growth of the RMI is clearly
mitigated when the thickness of the transition layer becomes
comparable to the modulation wavelength.  
Here we will consider the physical basis for this outcome. 

The competition of two timescales reasonably evaluates the effect of
the transition layer.  
One of those timescales is the transit time $\tau_{\rm tr}$ of the
incident shock to pass through the transition layer. 
It is given by $\tau_{\rm tr} \equiv L / \langle U_i \rangle$ where
$\langle U_i \rangle$ is the averaged shock velocity in the transition
layer. 
The other one is the stabilizing time $\tau_{\rm st}$ for the pressure
fluctuations by sound waves, which is expressed as $\tau_{\rm st}
\equiv \lambda / \langle c^{\ast} \rangle$. 
Here $\langle c^{\ast} \rangle$ is the averaged sound speed at the
downstream of the shock. 


When the transit time is shorter than the stabilizing time, the
transition layer has little effect on the growth of the RMI. 
However, if $\tau_{\rm st} \lesssim \tau_{\rm tr}$, the RMI growth
should be modified by the presence of the transition layer.  
Then the suppression condition is given by $L \gtrsim \zeta \lambda$,
where $\zeta \equiv \langle U_i \rangle / \langle c^{\ast} \rangle$.
We can guess the size of $\zeta$ assuming $\langle U_i \rangle \sim M
c_{b0}$ and $\langle c^{\ast} \rangle \sim ( c_{a}^{\ast} +
c_{b}^{\ast} ) / 2$, where $c_{a}^{\ast}$ ($c_{b}^{\ast}$) is the
sound speed of the post-shocked fluid ``a'' (``b'') at the interface
for the $L = 0$ case.  
It turns out by solving appropriate Riemann problems that $\zeta$
is of the order of unity for most of the cases we examined 
(see Table~\ref{tab1}).  
Therefore the suppression condition is approximately given by
\begin{equation}
  \frac{\tau_{\rm tr}}{\tau_{\rm st}} \sim \frac{L}{\lambda} \gtrsim 1
\end{equation}  
which is consistent with our numerical results.
Interestingly, this interpretation is independent of the density
gradient, which brings a difference from the stability condition for the
Kelvin-Helmholtz instability \cite{chandrasekhar61}.   

If the transition layer is broader than the wavelength of the
interface modulation,
the kinetic energy of the RMI motions
decreases with the power of $L^{-p}$ where $p \sim 2.5$. 
This result is equivalent that the growth velocity of the RMI is
reduced in proportion to $L^{-p/2}$ when $L \gtrsim \lambda$. 
There might be several reasons for the suppression of the RMI due to
the density transition layer. 


\begin{figure}
\includegraphics[scale=0.85,clip]{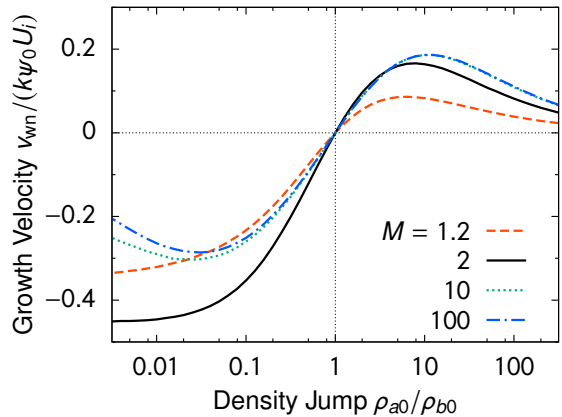}%
\caption{
The linear growth velocity of the WN formula calculated from
Eq.~(\ref{eq:dvi}) shown as a function of the density jump $\rho_{a0} /
\rho_{b0}$ for various Mach number cases; $M = 1.2$ (red dashed), $2$
(black solid), $10$ (green dotted), and $100$ (blue dot-dashed).
The growth velocity is normalized by $k \psi_0 U_i$.
For the isentropic exponent, $\gamma_{a} = \gamma_{b} = 5/3$ is
assumed in all cases.
\label{fig5}}
\end{figure}

If $\tau_{\rm st}$ is shorter than $\tau_{\rm tr}$, the density
difference felt by the incident shock is smaller than the
discontinuous case, $\Delta \rho_0 \equiv |\rho_{a0} - \rho_{b0}|$. 
The effective density difference is then estimated as $\Delta
\rho_{\rm eff} \sim \Delta \rho_0 \lambda / L$ assuming the linear
density gradient.
In a range of $0.01 \le \rho_{a0}/\rho_{b0} \le 100$ examined in our
analysis, the growth velocity has a complicated dependence on the
density jump \cite{wouchuk15}. 
Figure~\ref{fig5} shows the growth velocity of the WN model as a
function of the density jump $\rho_{a0} / \rho_{b0}$ for different
Mach number cases.
In this figure, we assume $\gamma_a = \gamma_b = 5/3$ for the
isentropic exponent.
For the case of $\gamma_a = \gamma_b$, the growth velocity of the RMI
must be zero when $\rho_{a0} / \rho_{b0} = 1$.

In the limit of the small density difference $\Delta \rho_0 /
\rho_{b0} \ll 1$, the asymptotic linear growth velocity has a scaling law of
the form 
\cite{coboscampos16,coboscampos17} 
\begin{equation}
  \frac{| v_{\rm wn} |}{k \psi_0 U_i} \approx
  c_1^{s,r} \frac{\Delta \rho_0}{\rho_{b0}}
  + O \left( \frac{\Delta \rho_0^2}{\rho_{b0}^2} \right) \;,
\label{eq:vlim}
\end{equation}
where a coefficient $c_1^{s,r} (> 0)$ has different expressions for shock- or
rarefaction-reflected cases (see Appendix). 
Replacing $\Delta \rho_0$ in Eq.~(\ref{eq:vlim}) with the effective
density difference $\Delta \rho_{\rm eff}$, the growth velocity in the
limit of $L / \lambda \gg 1$ is obtained by
\begin{equation}
\frac{| v_{\rm wn} |}{c_{b0}} \sim
2 \pi c_1^{s,r} M \frac{\Delta \rho_0}{\rho_{b0}} 
\left( \frac{\psi_0}{\lambda} \right)
\left( \frac{L}{\lambda} \right)^{-1} \;.
\end{equation}
This relation suggests that the mitigation of the RMI is larger as the
transition layer 
becomes broader, and which implies the qualitative coincidence with
the numerical results shown by Fig.~\ref{fig4}.  

The physical reason behind the RMI suppression is that as the
transmitted shock advances through the transition layer, its ripple
decreases.
In the RMI, perturbations are generated as the result of the conservation of the tangential momentum across the fronts.
Thus, the smaller the shock ripple is, the weaker the perturbations are.
In the end,
the mixing motions
developed by the RMI is weakened due to the transition layer.
Likewise, weakened shock strength because of the smooth density
gradient promotes the suppression and affects the quantitative
dependence of the growth velocity.
Thus, the index $p \sim 2.5$ might be determined by the combination of
multiple origins, although the value seems to be valid in a wide range
of parameters.
Analytic study on the transition-layer effects would be challenging
future work but inevitable for further understanding. 

In this work, we assume the isentropic exponent is constant everywhere $\gamma_a = \gamma_b = 5/3$.
The suppression due to the transition layer is affected by $\gamma$ through the stabilizing time $\tau_{\rm st}$.
Then, for a given thickness of $L$, the stiffer equation of state
would be easier to reduce the perturbed velocity of the RMI.

Lastly, we consider the application of our results for laboratory laser
plasmas.
The existence of a laser ablation plasma at a target surface could
play a role as a transition layer during shock interaction.
Exponential distribution of the density is often assumed for the
ablation plasmas. 
Suppose a case of $\rho_{a0}/\rho_{b0}=0.01$, for instance, we need to
decide the interface density $\rho_{ai}$ as for the edge value of the
exponential distribution. 
The interface density $\rho_{ai}$ would depend on the details
of the target density, laser intensity, and pulse shape so that it
has substantial ambiguity.  


\begin{figure}
\includegraphics[scale=0.85,clip]{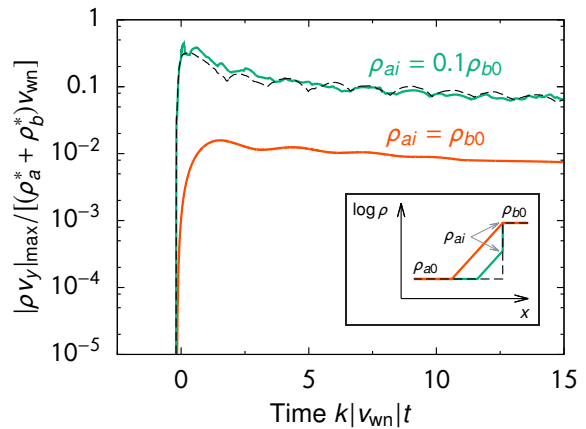}%
\caption{
Time evolutions of the maximum tangential velocity in the exponential
transition layer for different interface density of $\rho_{ai} =
\rho_{b0}$ (red) and $\rho_{ai} = 0.1 \rho_{b0}$ (green).
The density jump is $\rho_{a0} / \rho_{b0} = 0.01$ and the scale
length is assumed as $L / \lambda = 10$ for both cases.  
The other parameters are the same as in the fiducial run ($M = 2$ and
$\psi_0 / \lambda = 0.1$).
The time profile in the corresponding discontinuous case of $L = 0$ is
also shown by the black dashed curve. 
(inset) Initial density profiles in the logarithmic scale for two cases
of $\rho_{ai} = \rho_{b0}$ (red) and $\rho_{ai} = 0.1 \rho_{b0}$
(green).
\label{fig6}}
\end{figure}

Then, we perform demonstrative calculations with different $\rho_{ai}$
in the suppression case of $L/\lambda = 10$, which is shown in
Fig.~\ref{fig6}.  
This figure shows the maximum of the tangential momentum $|\rho v_y|$
searched from the entire domain for a given time.
The behaviors of the RMI are profoundly affected by the assumption of
$\rho_{ai}$. 
Even if ten percent of the density jump exists, i.e., $\rho_{ai} = 0.1
\rho_{b0}$, the RMI growth in the $L/\lambda = 10$ case becomes as
active as the no-transition case. 
In order to benefit from the stabilization by the
transition layer, the density distribution should be continuous from
$\rho_{b0}$ to $\rho_{a0}$. 
This could be crucial in designing laser experiments for inertial
confinement fusion, where the elimination of the interfacial
instabilities is really demanded \cite{atzeni04,betti16}. 

\section{Conclusions \label{sec5}}

We have investigated the role of the density transition layer on the
growth of the RMI
using two-dimensional hydrodynamic simulations.  
{\color{black}{
Although three-dimensional evolutions of hydrodynamic instabilities 
are essential in many cases, the two-dimensional study is still
important to understand the physics behind it.}}
A universal condition for the suppression of RMI due to the transition
layer has been obtained successfully through the systematic parameter study.  
If the transition layer is narrower than the wavelength of the surface
modulation, the effect on the RMI is ignorable.  
However,
the RMI growth
is severely reduced when the thickness
of the transition layer exceeds the modulation wavelength. 
The obtained threshold condition, $L \gtrsim \lambda$, can be
explained by the comparison between the shock-transit time through the
transition layer and the stabilizing time of the pressure fluctuations
by sound waves. 
This simple criterion will be useful to evaluate the importance of the
RMI in various situations, such as interstellar shock waves in
astrophysical phenomena and laser-driven shocks in inertial
confinement fusion experiments. 

An exhaustive analytic study on the transition-layer effects should be
necessary. Nonetheless, the inclusion of compressibility effects and
double reflection of reflected waves makes the calculations extremely
cumbersome, and it is proposed as future work.

\begin{acknowledgments}
We thank K. Mima, K. Nishihara, H. Sakagami, Y. Sentoku and
J. G. Wouchuk for useful discussions and encouragement.
Computations were carried out on SX-ACE Lite at the Institute of Laser
Engineering, Osaka University.  
This work was partly performed under the joint research project of the
Institute of Laser Engineering, Osaka University.  
This research was supported by JSPS KAKENHI Grant No. JP26287147 and
No. JP19KK0072, JSPS Core-to-Core Program, B. Asia-Africa Science
Platforms No. JPJSCCB20190003, and MEXT Quantum Leap Flagship Program Grant
No. JPMXS0118067246.
F.C.-C. has received support from MINECO under Grant
No. ENE2016-75703-R, from JCCM Grant No. SBPLY/17/180501/000264,
and from BBVA Foundation Leonardo Grant No. 2019/00570/001.
\end{acknowledgments}

\appendix*
\section{The growth velocity of the RMI in the limit of small density jump}

\begin{figure}
\includegraphics[scale=0.85,clip]{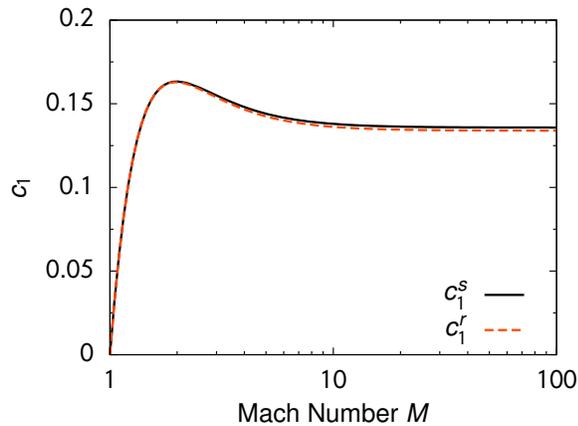}%
\caption{
Mach number dependence of the coefficients $c_1^{s}$ (black solid) and
$c_1^{r}$ (red dashed) given by Eqs.~(\ref{c1s}) and (\ref{c1r}) for
the case of $\gamma = 5/3$. 
\label{fig7}}
\end{figure}

In the limit of small pre-shocked density jump, the asymptotic growth
velocity $v_{\rm wn}$ is given by the scaling laws provided by Eq. (56) in
\cite{coboscampos16} and Eq. (167) in \cite{coboscampos17}, for the
shock- and rarefaction-reflected cases, respectively.
Assuming $\gamma_a=\gamma_b=\gamma$, the constant term $c_0^{s,r}$ of the
scaling laws becomes zero, and, hence, the growth velocity is
proportional to the density difference as indicated in Eq. (\ref{eq:vlim}). 
The first term coefficients $c_1^s$ and $c_1^r$ are approximately
given by the following expressions as a function of the Mach number
$M$ and the isentropic exponent $\gamma$.

\begin{widetext}
For the shock-reflected case:
\begin{eqnarray}
c_1^s (M,\gamma) &=& - \frac{\pi_3^2 (M^2-1)}
{(\pi_1 + 2\pi_2\pi_3\pi_4) (\gamma+1)^2M^2 [2(\gamma-2)M^2+\gamma-1]
  [(2\gamma-1)M^4+2M^2+1]} \nonumber \\ 
&\times &  \left( \pi_1\left[ \pi_3\pi_4 (3M^2+1)+2(-2\gamma^2+2\gamma+1)M^4+(-2\gamma^2-3\gamma+3)M^2-\gamma+1\right] \right.  \nonumber \\
&+& \pi_2\left\{ \pi_3\pi_4 \left[
  (-8\gamma^2+7\gamma+3)M^4-4(\gamma^2+\gamma-2)M^2-3\gamma+1\right] 
\right. \nonumber \\
&+&\left. \left. 2(6\gamma^2-5\gamma+1)M^6 + (-\gamma^2+28\gamma-11)M^4+2(-2\gamma^2+\gamma+7)M^2+\gamma^2-4\gamma+3\right\}
\right) \;, 
\label{c1s}
\end{eqnarray}
where
\begin{eqnarray}
\pi_1 &=& \left[ ( 9\gamma^3 - 13\gamma^2 + 11\gamma + 1 ) M^6 + (
  -7\gamma^3 + 35\gamma^2 - 53\gamma + 1 ) M^4 \right. \nonumber \\ 
&+& \left.( 3\gamma^3 - 7\gamma^2 + 73\gamma - 13 ) M^2 - \gamma^3 -
  3\gamma^2 - 19\gamma + 15 \right]^{1/2} \; , \\ 
\pi_2 &=& \left[ ( \gamma + 1 )( M^2 - 1 ) \right]^{1/2}  \; , \\ 
\pi_3 &=& \left[ 2+(\gamma-1)M^2 \right]^{1/2}  \; , \\ 
\pi_4 &=& \left[ 1+\gamma(2M^2-1) \right]^{1/2}  \; . 
\label{dt_1}
\end{eqnarray}

For the rarefaction-reflected case:
\begin{equation}
c_1^r(M,\gamma)=\frac{2}{(\gamma+1)^2M^2}\frac{\gamma(\gamma-1)M^6-(\gamma^2-4\gamma+1)M^4-3(\gamma-1)M^2-2}{(2\gamma-1)M^4+2M^2+1} \;,
\label{c1r}
\end{equation}
which is valid when $1\leq \gamma \leq 3$.
For $\gamma>3$, the expression is very cumbersome and impractical to use.
Therefore, we decide not to show here, considering that the cases with
$\gamma>3$ are quite rare. 

The dependence of $c_1^{s,r}$ on the Mach number calculated by
Eqs.~(\ref{c1s}) and (\ref{c1r}) is shown by
Fig.~\ref{fig7} for the case of $\gamma = 5/3$.
These two coefficients take similar values for this case, although the
formulas are quite different.
The behavior of $c_1^{s,r}$ is consistent with the growth velocity of
the WN model around $\rho_{a0} / \rho_{b0} = 1$ (see Fig.~\ref{fig5}).
\end{widetext}

%


\end{document}